\documentstyle[preprint,aps]{revtex}
\begin{document}
\draft
\title{The nature and validity of the RKKY limit of exchange coupling in
magnetic trilayers}
\author{M. S. Ferreira \dag, R. B. Muniz \ddag,
J. d'Albuquerque e Castro\thanks{Permanent address: Instituto de F\'{\i}sica,
Universidade Federal Fluminense.} \dag\dag\hspace{0.1cm}  and D. M. Edwards
\dag}
\address{\dag  Department of Mathematics, Imperial College, London, SW7 2AZ, UK
\\
\dag\dag  City University, London, EC1V 0HB, UK \\
\ddag  Instituto de F\'{\i}sica, Universidade Federal Fluminense, Niter\'oi,
Rio de Janeiro, 24001-970, Brazil}
\date{\today}
\maketitle
\begin{abstract}
The effects on the exchange coupling in magnetic trilayers due to the presence
of
a spin-independent potential well are investigated. It is shown that within the
RKKY
theory no bias nor extra periods of oscillation associated with the depth of
the
well are found, contrary to what has been claimed in recent works. The range of
validity of the RKKY theory is also discussed.
\end{abstract}
\vspace{1cm}

% body of paper here

Metallic magnetic multilayers are systems composed of alternating ultrathin
layers of magnetic and non-magnetic metallic materials.
They exhibit exchange coupling between magnetic layers
across a non-magnetic spacer. Such a coupling oscillates between ferromagnetic
and antiferromagnetic and its strenght decays as the spacer layer thickness is
varied.

Basically, two mechanisms have been proposed for explaining the oscillatory
coupling in multilayers. Edwards et al. \cite{r4,r5} have related the
phenomenon
to
the existence of quantum wells for electrons (or holes) with both
spin orientations propagating through the multilayer structure, which result
from the
exchange interaction inside the magnetic layers. The existence of
those quantum wells was lately confirmed by photoemission experiments
\cite{r6}.
According to the model proposed by Edwards et al., the observed oscillation
in the coupling results from quantum interference effects inside the well
and bears a formal analogy to the de Haas-van Alphen effect.
The other proposed mechanism for the coupling in multilayers is due
to Bruno and Chappert \cite{r7} and is based on an extension of the RKKY theory
to the particular geometry of the system. In their approach the
coupling arises from the polarization of magnetic carriers in the spacer,
as in the ordinary RKKY theory for the coupling between localized magnetic
moments in a non-magnetic host.

The formal connection between the two theories was recently made explicit
by d'Albu-querque e Castro et al \cite{r8}. In their work, model-independent
closed form expressions are derived for the coupling, as well as for
the bilinear and biquadratic terms, and it is shown that the full
confinement and the RKKY-like  theories correspond to limiting cases in
which the exchange interaction inside the magnetic layers is very strong
and very weak, respectively. Another important point made by
d'Albuquerque e Castro at al. is that in the usual RKKY-like theories
the difference between the non-magnetic parts of the potentials in the
spacer and in the magnetic layers are neglected, which may lead to
significant errors in the amplitude and phase of the coupling.

Both the quantum well and RKKY-like theories relate the period of
oscillation of the coupling  to the extremal points of the spacer
Fermi surface in the direction perpendicular to the layers. However,
in a recent paper by Jones et al. \cite{r9} the question
of other possible sources for the period of oscillation is discussed and it is
claimed
that the oscillation period is determined by the depth of the quantum well
formed by the
magnetic and spacer layers rather than by the dimensions of the spacer Fermi
surface.
In addition those authors
claim that such effects can generate a bias in the coupling, producing
a prefered ferromagnetic coupling. In recent calculations Mu\~noz and
P\'erez-D\'{\i}az
\cite{r10} have also
found periods in the coupling not related to the spacer Fermi surface, which
they claim are associated with the confined states in the quantum well.

In this communication we present results for the coupling
for cases in which the quantum well due to the difference in the potentials
of the spacer and the magnetic layers give rise to bound states for carriers
with both
spin directions. Those are just the situations
considered by Jones et al. and Mu\~noz   and P\'erez-D\'{\i}az. We have found
that,
provided the contributions to the coupling coming from singularities
associated with the presence of the bound states are properly dealt with,
no extra period nor any bias are obtained. We therefore conclude that
the features observed by Jones et al. and by Mu\~noz and P\'erez-D\'{\i}az are
spurious.

We consider trilayer systems described by the one-band tight-binding
model with nearest-neighbour hopping $t$ on a simple cubic lattice. The
layers are displaced perpendicularly to the $(100)$ direction and the site
energies are chosen equal to $6|t|$ in the spacer and $6|t|+V$ in the magnetic
material. Such a choice of parameters places the bottom of the spacer
band at the origin of the energies. For sufficiently small values of
the potential barrier $V$ and of the Fermi energy $E_{F}$, the Fermi surfaces
of both the spacer and magnetic materials are nearly spherical. Therefore
the effective mass approximation can be safely applied to both materials,
and the present model becomes equivalent to the electron gas model.
We follow Jones et al. and introduce a local exchange interaction $V_{ex}$
only in the magnetic atomic layers next to the spacer. We label these
two layers $0$ and $n$, so that the number of atomic layers in the spacer
is equal to $N=n-1$. The restriction of the local exchange interaction to just
one atomic layer in the magnetic material would have to be lifted if we were
to study the dependence of the coupling on the magnetic
layer thickness. A recent thorough investigation
of such a dependence \cite{r11}  has shown that as far as the behaviour of the
coupling as a function of the spacer thickness is concerned, changes in
the magnetic layer thickness just leads to changes in the phase and amplitude
of the coupling, keeping the period of oscillation unchanged. Therefore,
for the purpose of the present work the restriction of $V_{ex}$ to just one
monolayer in the magnetic material does not pose any limitation on the
validity of our final conclusions.

The change in the thermodynamical potential $\Omega$ at $T=0$ due to a rotation
by an angle $\theta$ of the magnetization in one of the magnetic layers
relative to that in the other layer is given by \cite{r8}

\begin{equation}
\Delta \Omega(\theta) \,=\,{1\over\pi} \,\sum_{\vec q_\parallel} \,
\int_{-\infty}^{E_f} d\,\omega\,Im\,ln\,\lbrace\,1 \,-\, 2 V^{\,2}_{ex} \,
(\,cos\,\theta -1\,) \,G^{\uparrow}_{n 0}(\omega,\vec q_\parallel)\,
G^{\downarrow}_{0 n}(\omega,\vec q_\parallel)\,\rbrace
\label{eq1}
\end{equation}

\noindent
where $G^{\sigma}_{0n}(\omega,\vec q_\parallel)$ is the off-diagonal
matrix element between sites $0$ and $n$ of the retarded Green's
function for an electron with spin $\sigma$ in the ferromagnetic $\theta=0$
configuration of the system. Here the summation over $\vec q_\parallel$
is restricted to the two-dimensional Brillouin zone. It is worth
stressing that the Green's function $G^{\sigma}$ properly takes into
account the presence of the potential well and, consequently, exhibits
poles at the corresponding bound states energies. Those poles lie on the
real axis. However, because the integrand in eq.(\ref{eq1}) is analytic in the
upper-half
of the complex energy plane, the integration
over the energy $\omega$ can be performed along a straight line joining the
points
$E_{F}+i \infty$ and $E_{F}$, thereby avoiding the singularities.

The expansion of the integrand in eq.(\ref{eq1}) in powers of $cos(\theta)$
\cite{r8}
enables us to introduce the bilinear $J_{1}$ and intrinsic biquadratic
$J_{2}$ coefficients, in terms of which the expression for $\Delta \Omega$
reads,

\begin{equation}
\Delta \Omega(\theta) = \Delta \Omega_0 \, - \, J_1 cos\theta \, - \, J_2
cos^2\theta
\label{eq2}
\end{equation}

\noindent
It is found that higher order terms than $cos^2\theta$ are negligible so that

\begin{equation}
J_{1} = \Delta \Omega(\pi)/2
\label{eq3}
\end{equation}

Here we are interested in those situations in which the local exchange
interaction $V_{ex}$ is small and the coupling can be calculated
by perturbation theory in that parameter. It is then straightforward
to show that to lowest order in $V_{ex}$ the bilinear coefficient
is given by \cite{r8}

\begin{equation}
J_{1}^{RKKY} =\,{1\over\pi} \,\sum_{\vec q_\parallel} \,
\int_{-\infty}^{E_f} d\,\omega\, Im\,\lbrace\, 2 V^{\,2}_{ex}\,
G^{\,0}_{n 0}(\omega,\vec q_\parallel)\,
G^{\,0}_{0 n}(\omega,\vec q_\parallel)\,\rbrace,
\label{eq4}
\end{equation}

\noindent
where the Green's functions are now calculated for $V_{ex}=0$. This
is just the second-order perturbation theory or RKKY result for the coupling.

Using the above expression we  evaluated $J_{1}^{RKKY}$ for the
trilayer system as a function of the spacer thickness $N$, for fixed values
of $E_{F}$ and $V_{ex}$, but for different well barriers $V$. For the systems
under
consideration it is possible to derive analytical expressions for the
off-diagonal
Green's function elements $G^{\,0}_{n 0}(\omega,\vec q_\parallel)$ and
$G^{\,0}_{0 n}(\omega,\vec q_\parallel)$ for an arbitrary value of
$n$\cite{r12}.
Those expressions can be analytically continued to non-integer $n$, which
permits us
to consider continuous variations of the spacer thickness. We set
$E_{F}=|t|$, which gives rise to a period of oscillation of exactly
three atomic layers, and $V_{ex}=0.04|t|$. In all cases, the calculated
values of $J_{1}^{RKKY}$ (diamonds) are compared, following eq.(\ref{eq3}),
with
$\Delta \Omega(\pi)/2$ (solid circles), the latter quantity being calculated
from
the full expression (\ref{eq1}).

Fig. (\ref{fig1}a) shows results for the case in which $V=0$ (no potential
well).
As expected the coupling oscillates about zero (no bias) and with a well
defined period of three atomic layers. However, in contrast with what
was found by Jones et al. and Mu\~noz and Perez-D\'{\i}az, those features
are not changed when non-zero values of $V$ are considered. We have
performed calculations for $V$ equal to $0.3E_{F}$, $0.6E_{F}$, and
$0.9E_{F}$. Results are shown in Figs. (\ref{fig1}b), (\ref{fig1}c), and
(\ref{fig1}d),
respectively.
The only noticable changes in the curves are in the phase and in the
amplitude of oscillations. We have observed, however, that because of the
singular behaviour of the integrand along the real energy axis, spurious
features, such as bias and extra period, can be easily obtained when the
contributions from the singularities are not properly taken into account.
We have explicitly verified this point by evaluating the energy integration
also
along the real axis. It should be stressed that numerical problems may
occur even when $V=0$. In such a case, for each $\vec q_\parallel$, the
integrand
in Eq.(4) has a
pole right at the bottom of the band. In all cases, if the contribution from
poles
is not treated properly,
a bias in the coupling is obtained. This point was already noticed by Yafet
\cite{r13}
in the case of the RKKY interaction for the one-dimensional electron gas with
$V=0$

It is interesting to notice that for the value of $V_{ex}$ considered
above, the second order perturbation result $J_{1}^{RKKY}$ for the
coupling agrees well with the full calculation $\Delta \Omega(\pi)/2$.
For higher values of this parameter, however, the second order perturbation
theory does not provide a satisfactory description of the coupling and full
calculations have to be performed. This point is illustrated in
Fig.(\ref{fig2}) which shows results of both $J_{1}^{RKKY}$ (diamonds) and
full calculation (solid circles) as functions of $V_{ex}$, for different
values of $V$ and $N$. We clearly see that for fixed $E_{F}$ the range
of applicability of the second order perturbation theory strongly depend
on both $N$ and $V$.

In summary, we have shown that the inclusion of interface effects through a
spin independent square well potential does not alter the period of
oscillation of the RKKY bilinear coupling, as it was recently claimed.
We have also checked that the appearance of a ferromagnetic bias is due
to numerical inaccuracy in the calculation and does not reflect any real
effect due to presence of bound states. Finally, we have explicitly shown
the limited validity of the RKKY-like theories.

Financial support from CNPq of Brazil is gratefully acknowledged.

% now the references. delete or change fake bibitem. delete next three
%   lines and directly read in your .bbl file if you use bibtex.

\begin{figure}
\caption{Exchange Coupling as a function of spacer thickness for barrier
heights $V = 0$ (a), $0.3E_{F}$ (b), $0.6E_{F}$ (c), and $0.9E_{F}$ (d). In
each
graph,
results for $J_{1}^{RKKY}$ are represented by diamonds and those for
$\Delta\Omega(\pi)/2$
by solid circles. Full lines are for continous values of n.}
\label{fig1}
\end{figure}
\begin{figure}
\caption{Exchange Coupling as a function of the local exchange interaction
$V_{ex}$ for
different
heights of the well and spacer thicknesses. Results for $J_{1}^{RKKY}$ are
represented
by diamonds and those for $\Delta\Omega(\pi)/2$
by solid circles.}
\label{fig2}
\end{figure}

% \begin{table}
% \caption{}
% \label{}
% \end{tabular}
% \end{table}

\end{document}